\documentclass[a4paper]{jpconf}
\usepackage{graphicx}
\usepackage{amssymb}
\usepackage{tabularx}
\begin{document}
\title{Dark energy in vector-tensor theories of gravity}

\author{Jose Beltr\'an Jim\'enez$^\dag$ and Antonio L. Maroto$^\ddag$}

\address{Dpto. F\'isica Te\'orica, Universidad Complutense de Madrid, 28040, Madrid, Spain.}

\ead{$\dag$ jobeltra@fis.ucm.es; $^\ddag$  maroto@fis.ucm.es}

\begin{abstract}
We consider a general class of vector-tensor theories of gravity and show that solutions with accelerated expansion and a future type III singularity are a common feature in these models.  We also show that there are only six vector-tensor theories with the same small scales behavior as General Relativity and, in addition, only two of them can be made completely free from instabilities. Finally, two particular models as candidates for dark energy are proposed: on one hand, a cosmic vector that allows to alleviate the usual naturalness and coincidence problems and, on the other hand, the electromagnetic field is shown to give rise to an effective cosmological constant on large scales whose value can be explained in terms of inflation at the electroweak scale.
\end{abstract}
\vspace{-0.14cm}
\section{Introduction}
The standard model of Cosmology accounts for the phase of accelerated expansion that the universe is currently undergoing by introducing a cosmological constant term in the gravitational action. According to observations, the value of this cosmological constant should be $(10^{-3}$ eV$)^4$, which is very far away from the natural scale of gravitation set by the Planck mass $M_P=1/\sqrt{G}\sim 10^{19}$ GeV. This fact poses the so-called naturalness problem associated to the cosmological constant. However, this could be signaling that the cosmological constant causing the accelerated expansion is not a true constant, but an effective description arising from some underlying physical theory. On the other hand, the accelerated expansion could also be caused by some other physical mechanism like a cosmological scalar field or  a modification of the gravitational interaction \cite{scalarfields}. Nonetheless, in most of these models, the original naturalness problem still remains because new dimensional scales are needed in order to produce the correct acceleration. Furthermore, they usually have some other problems related to the presence of instabilities or local gravity constrains. In this work we shall consider vector-tensor theories of gravity and show how vector fields can help to alleviate the naturalness problem. Indeed, we shall show that the very familiar electromagnetic field can play the role of dark energy without any of the aforementioned problems.

\section{Vector-tensor theories of gravity}
We shall consider the most general action for a vector-tensor theory leading to second order linear equations of motion for the vector field:
\begin{eqnarray}
S[g_{\mu\nu},A_\mu]=\int d^4x\sqrt{-g}\left[-\frac{1}{16\pi
G}R+\omega RA_\mu A^\mu+\sigma R_{\mu\nu}A^\mu
A^\nu
+\lambda(\nabla_\mu A^\mu)^2+\epsilon F_{\mu
\nu}F^{\mu\nu}\right]\label{action}
\end{eqnarray}
with $\omega$, $\sigma$, $\lambda$ and $\epsilon$ dimensionless parameters. The interest of this theory is that $G$ is the only dimensional constant and we do not have potential terms for the vector field so that no new scales are introduced in the action. A general study of the homogeneous cosmological evolution for this action is given in \cite{Jimenez:2009ai} where a classification attending to the behavior of the vector field throughout the different epochs of the universe evolution is performed. Also, the case in which the vector field dominates the universe is considered and the corresponding autonomous system is analyzed. For this system, it is shown the presence of both attractors and repellers in which accelerated expansion can take place. These repellers are interesting for inflaton models because we have an initial state with accelerated expansion that can eventually end as the trajectory in the phase space goes away from the critical point. On the other hand, attractors with accelerated expansion can give rise to dark energy candidates since they provide late-time accelerated expansion. Finally, when a matter component is introduced in addition to the vector field, we have that, for quite general conditions, there are solutions with a transition from a matter dominated universe to a phase of accelerated expansion with a phantom behavior for the vector field that leads to a future type III singularity in which the scale factor remains finite but the expansion rate diverges.

The small scales behavior of the vector-tensor theory given by (\ref{action}) is well described by means of its PPN parameters. Since GR is in excellent agreement with Solar System experiments, an alternative gravitational theory whose PPN parameters are the same as those of GR will agree with small scales viability conditions. The PPN parameters are obtained by performing perturbations around a Minkowski spacetime with a constant background vector field of the form $A_\mu=(A,0,0,0)$. When we impose the PPN parameters of the vector-tensor theory to be identical to those of GR we find that there are six models which satisfy such a condition  for any value of the background vector field \cite{Jimenez:2008sq}. These six models have $\omega=0$ and can be classified as: Gauge non-invariant models:   $i)\;
\sigma=-4\lambda=-4\epsilon$, $ii)\;\sigma=-3\lambda=-2\epsilon$, $iii)\;\sigma=0$; Gauge invariant models: $\lambda=0$ and $\sigma=m\epsilon$ with $m=0,-2,-4$. For these particular vector-tensor theories we study the presence of instabilities both at the classical and the quantum level in the inhomogeneous perturbations\footnote{We consider perturbations in both the vector field and the metric tensor.}. The classical stability can be achieved by imposing the absence of modes whose propagation speed is imaginary so that we do not have exponentially growing modes. On the other hand, the quantum stability will be guaranteed by eliminating the modes with negative energy (ghosts) in order to have a stable quantum vacuum (see Table 1). We find that the only two theories which can be made completely free from instabilities are the pure Maxwell action and Maxwell action supplemented with a gauge fixing term.

\begin{table}[h!]
\begin{center}
{\scriptsize \hspace{0.3cm}
\begin{tabularx}{15.9cm}{|X|c|c|c|c|} \hline
& & & &\\
& Model I  & Model II  & Model III & Gauge invariant models \\
&   &   &  & $m=-2,\, -4$ \\
\hline & & & & \\Classical stability & $-32\epsilon A^2<1$ &
 $-15\lambda A^2<1$ &Always & $
\epsilon
A^2\notin\left[\frac{1}{\left(2+\frac{1}{2}m\right)m},\frac{1}{2m}\right]
$
\\& & & & \\
\hline & & & &\\ Gravitational waves & $-16\epsilon A^2<1$ &
 $-12\lambda A^2<1$ & Unaffected & $4m\epsilon A^2<1$
\\& & && \\
\hline & & & &\\ Quantum stability & $1<-64\epsilon A^2<4$ &
 $\lambda A^2\in(-0.098,-0.033)$ & $\partial_\mu A^\mu=0$ and $\epsilon<0$ &
 $\varepsilon A^2\in\left(a_-,\frac{1}{2m}\right)
\cup\left(b_+,0\right)$\\
& & & & \\
\hline  & & & &\\Viability condition & $1<-64\epsilon A^2<2$ &
$\lambda A^2\in\left(-\frac{1}{15},-0.033\right)$ &$\partial_\mu
A^\mu=0$ and $\epsilon <0$ & $\epsilon
A^2\in\left(b_+,\frac{1}{4m}\right)$
\\ & & & &\\
\hline
\end{tabularx}
\label{tablaPPN}\vspace{0.1 cm}} \caption{\small In this table we
summarize the conditions obtained in order to have both classical
and quantum stability for the models with the same set of PPN
parameters as GR. The $m=0$ gauge invariant
model satisfies all the viability conditions.}
\end{center}
\end{table}

\section{Scaling vector dark energy}
The cosmic vector model having a scaling behavior in the early universe proposed in \cite{Jimenez:2008au} is described by the following action:
\begin{equation}
S=\int d^4x\left[-\frac{R}{16\pi G}-\frac{1}{4}F_{\mu\nu}F^{\mu\nu}-\frac{1}{2}(\nabla_\mu A^\mu)^2+R_{\mu\nu}A^\mu A^\nu\right].
\end{equation}
In this model, the fraction of energy density stored in the vector field remains constant throughout the radiation dominated epoch and, in addition, the value of the vector field  during such an epoch also remains constant. This prevents any dependence on the moment at which we set the initial conditions because the subsequent evolution of the universe will be the same irrespective of the initial redshift (as long as it is well inside the radiation dominated epoch). Moreover, the initial fraction of energy density corresponding to the vector field needed to explain the SNIa measurements turns out to be $\sim 10^{-6}$ and the vector field must take a value $\sim 10^{-4} M_P$, which are values that can arise {\it naturally} in the early universe, as quantum fluctuations for instance.  That way, this model allows to alleviate the naturalness problem and it does not require any fine-tunings.

Finally, we have confronted the predictions of the model to SNIa, CMB and BAO measurements. An interesting property of this model is that it provides the best fit to date to the SN Gold dataset with the same number of parameters as the standard $\Lambda$CDM. Moreover, CMB measurements predicts a closed universe for this model, being the flat case strongly disfavored. Finally, we find some tension with BAO observations, which could be due to the dependence of such data on the fiducial model ($\Lambda$CDM) whose evolution is very different from that given by the cosmic vector model.

\section{Electromagnetic dark energy}
The second vector dark energy model we are going to discuss is nothing but the very familiar electromagnetic field. It is well-known \cite{Itzykson} that the canonical quantization of the electromagnetic field needs to impose the subsidiary Lorenz condition $\nabla_\mu A^\mu =0$ in order to reduce the corresponding Hilbert space so that we end up with the two transverse states for the photon.  However, when we consider the covariant quantization in an expanding universe, the Lorenz condition turns out to be difficult to implement. This is due to the fact that $\nabla_\mu A^\mu$ behaves as a free scalar field so that it can be excited by means of gravitational fields. For these reasons,  we consider the action including a gauge fixing term\cite{Jimenez:2009dt}:
\begin{equation}
S=\int d^4x\sqrt{-g}\left[-\frac{1}{4}F_{\mu\nu}F^{\mu\nu}+\frac{\xi}{2}(\nabla_\mu A^\mu)^2+A_\mu J^\mu\right]\label{EMaction}
\end{equation}
as the fundamental action for the electromagnetic field. In this action the gauge fixing term is a physical term, rather than being a mathematical trick, so that we break the $U(1)$ invariance, although a residual gauge symmetry $A_\mu\rightarrow A_\mu+\partial_\mu\vartheta$ with $\Box \vartheta =0$ remains. Notice that this action is also the starting action in the path integral formalism once we have fixed the element of the orbits of the group. The corresponding modified Maxwell equations deduced from (\ref{EMaction}) are:
\begin{equation}
\nabla_\nu F^{\mu\nu}+\xi\nabla^\mu(\nabla_\nu A^\nu)=J^\mu.
\end{equation}
Now, if we take the four divergence of this equation we obtain:
$\Box(\nabla_\nu A^\nu)=0$
where we have used the conservation of the electromagnetic current. Thus, as we said before, $\nabla_\mu A^\mu$ behaves as a free scalar field non-minimally coupled to gravity. Moreover, in an expanding universe, a free scalar field gets frozen for super-Hubble modes so that the gauge fixing term appearing in (\ref{EMaction}) will give an effective cosmological constant on large scales. On the other hand, we also know that a free scalar field decays as $1/a$ for sub-Hubble modes which implies that the gauge fixing term will be negligible at small scales and we effectively recover the usual Maxwell equations.

Since we are breaking the $U(1)$ invariance and we do not impose any extra subsidiary condition, the theory contains one extra degree of freedom so that the field can be decomposed as follows:
$A_\mu=A_\mu^{(1)}+A_\mu^{ (2)}+A_\mu^{(s)}+\partial_\mu \vartheta$
where $A_\mu^{(i)}$ with $i=1,2$ are the two transverse modes of
the massless photon, $A_\mu^{(s)}$ is a new scalar state that
represents the mode that would have been eliminated if we had
imposed the Lorenz condition and, finally, $\partial_\mu \theta$
is a purely residual gauge mode, which can be eliminated by means
of a residual gauge transformation in the asymptotically free
regions, in a completely analogous way to the elimination of the
$A_0$ component in the Coulomb gauge.
The complete quantization procedure in an inflationary phase with De-Sitter expansion can be found in \cite{Jimenez:2009dt} where the primordial power spectrum defined by
$\langle 0\vert (\nabla_\mu A^\mu)^2\vert 0 \rangle =\int\frac{dk}{k}P_A(k)$ for super-Hubble modes is computed and given by:
\begin{equation}
P_A (k)=\frac{9 H_I^4}{16\pi^2}
\end{equation}
with $H_I$ the constant Hubble parameter during inflation. This result implies that the effective cosmological constant given by the gauge fixing term is $\sim H_I^4$ so that the measured value of the cosmological constant requires $H_I \sim 10^{-3}$ eV, which corresponds to an inflationary scale $M_I\sim 1$ TeV. Thus, we see that the cosmological constant scale can be naturally explained in terms of physics at the electroweak scale.

Although the homogeneous evolution is the same as in the standard $\Lambda$CDM, we have a crucial difference because, in this case, we can have inhomogeneous perturbations, unlike in the standard model where the cosmological constant is a true constant. In \cite{Jimenez:2009sv} we have considered the effects of the inhomogeneous perturbations on the CMB anisotropies and large scale structures and found that only at the largest scales we can have discriminatory consequences.   If the primordial amplitude of the perturbations are not extremely large, the CMB temperature and matter power spectra are perfectly compatible with observations.
\vspace{-0.3cm}
\section{Conlcusions}
We have shown that vector fields are compelling candidates to drive an accelerated phase and they can alleviate or even solve some of the usual naturalness or coincidence problems arising in most of dark energy models. More interestingly, we have shown that the electromagnetic field can account for the present phase of accelerated expansion of the universe and provides a natural explanation for the observed value of the cosmological constant. This shows that it is possible to establish the fundamental nature of dark energy without resorting to new physics.

\vspace{0.1cm}
{\em Acknowledgments}: This work has been supported by Ministerio de Ciencia e Innovaci\'on (Spain) project numbers FIS 2008-01323 and FPA 2008-00592, UCM-Santander PR34/07-15875, CAM/UCM 910309 and MEC grant BES- 2006-12059.
\vspace{-0.2cm}
\section*{References}
\providecommand{\newblock}{}

\end{document}